\documentclass[12,preprint]{aastex}
\newcommand{\kms}{\mbox{km s$^{-1}$}}

\newcommand{\Msun}{\mbox{$M_{\sun}$}}
\newcommand{\Lsun}{\mbox{$L_{\sun}$}}
\newcommand{\skipthis}[1]{}
\newcommand{\hii}{\mbox{\ion{H}{2}}}

\shortauthors{Garay et al.}
\shorttitle{High mass protostellar object}
\slugcomment{submitted to {\em The Astrophysical Journal}}
  
\begin{document}

%Draft \today

%%%%%%%%%%%%%%%%%%%%%%%%%%%%%%%%%%%%%%%%%%%%%%%%%%%%%%%%%%%%%%%%%%%%%%%%%
% TITLE and AUTHORS
%%%%%%%%%%%%%%%%%%%%%%%%%%%%%%%%%%%%%%%%%%%%%%%%%%%%%%%%%%%%%%%%%%%%%%%%%

\title{Two massive star forming regions at early evolutionary stages}
  
\author{Guido Garay}
\affil{Departamento de Astronom\'{\i}a, Universidad de Chile,
Casilla 36-D, Santiago, Chile}
% guido@das.uchile.cl

\author{Kate Brooks}
\affil{European Southern Observatory, Casilla 19001, Santiago 19, Chile}

\author{Diego Mardones}
\affil{Departamento de Astronom\'{\i}a, Universidad de Chile,
Casilla 36-D, Santiago, Chile}
% mardones@das.uchile.cl

\author{Ray P. Norris}
\affil{Australia Telescope National Facility, P.O. Box 76, Epping 1710 NSW, Australia}

\and

\author{Michael G. Burton}
\affil{Department of Astrophysics and Optics, School of Physics, University of New 
South Wales, NSW 2052, Australia}

%%%%%%%%%%%%%%%%%%%%%%%%%%%%%%%%%%%%%%%%%%%%%%%%%%%%%%%%%%%%%%%%%%%%%%%%%
% Abstract
%%%%%%%%%%%%%%%%%%%%%%%%%%%%%%%%%%%%%%%%%%%%%%%%%%%%%%%%%%%%%%%%%%%%%%%%%
\begin{abstract}

We report sensitive ATCA radio continuum observations toward IRAS 15596$-$5301 
and 16272$-$4837, two luminous objects (${\cal L}>2\times10^4$\Lsun) thought to 
represent massive star forming regions in early stages of evolution (due to 
previously undetected radio emission at the 1$\sigma$ level of 2 mJy per beam). Also 
reported are 1.2 millimeter continuum and a series of molecular line observations made 
with the SEST telescope.

The radio continuum observations toward IRAS 15596$-$5301 reveal the presence of three 
distinct compact sources, with angular sizes of 2.7\arcsec\ to 8.8\arcsec\ (FWHM), all 
located within a region of 30\arcsec\ in diameter. Assuming that these are regions of 
ionized gas, we find that they have diameters of 0.06-0.2 pc, electron densities of 
$8\times10^2 - 2\times10^3$ cm$^{-3}$, and that they are excited by early B type stars. 
The 1.2-mm observations show that the dust emission arises from a region of 
$42\arcsec\times25\arcsec$ (FWHM) with a total flux of 5.8 Jy, implying a mass of 
$1.4\times10^3$ \Msun. The line observations indicate that IRAS 15596$-$5301 is 
associated with a molecular cloud with a FWHM angular size of 37\arcsec\ ($\sim0.4$ pc 
radius at the distance of 4.6 kpc), a molecular hydrogen density of $\sim4\times10^5$ 
cm$^{-3}$ and a rotational temperature of $\sim27$ K. We suggest that the massive 
dense core associated with IRAS 15596$-$5301 contains a cluster of B stars which are 
exciting compact \hii\ regions that are in pressure equilibrium with the dense 
molecular surroundings. 

No radio continuum emission was detected from IRAS 16272$-$4837 up to a $3\sigma$ 
limit of 0.2 mJy. However, the 1.2-mm observations show strong dust emission 
arising from a region of $41\arcsec\times25\arcsec$ (FWHM) with a total flux of 13.8 
Jy, implying a mass of $2.0\times10^3$ \Msun. The line observations indicate the 
presence of an elongated molecular cloud with FWHM major and minor axes of 61\arcsec\ 
and 42\arcsec ($0.50\times0.35$ pc in radius at the distance of 3.4 kpc), a 
molecular hydrogen density of $\sim2\times10^5$ cm$^{-3}$ and a rotational temperature 
of $\sim27$ K. The high luminosity ($2.4\times10^4$\Lsun) and lack of radio emission 
from this massive core suggest that it hosts an embedded young massive protostar that 
is still undergoing an intense accretion phase. This scenario is supported by the 
observed characteristics of the line profiles and the presence of a bipolar outflow 
detected from observations of the SiO emission. We suggest that IRAS 16272$-$4837 is 
a bona-fide massive star forming region in a very early evolutionary stage, being the 
precursor of an ultra compact \hii\ region. 

\end{abstract}


\section{Introduction}

The earliest phase of high-mass star formation is possibly one of the least 
understood stage of evolution of massive stars. Massive stars (M$>8$ \Msun) are known 
to be formed in dense molecular cores, however the sequence of processes leading to 
their formation is not yet well established. In particular, the role of coalescence 
(Stahler et al. 2000) and accretion (Osorio, Lizano, \& D'Alessio 2000) processes in 
the assembling of a massive star is still under debate. The determination of the 
physical conditions of the gas during the formation and early evolution of a massive 
star is difficult because of their scarcity and rapid evolution. In addition, massive 
stars are usually born in clusters or groups hence their individual studies are usually 
afflicted by confusion, particularly because they are found located, on the average, 
at larger distances from the Sun than sites of low-mass star formation.
Massive objects in early evolutionary stages, namely in the process of assembling 
the bulk of their final mass, should be identified by having high bolometric 
luminosities ($>10^4 $\Lsun), strong dust emission, and very weak or no detectable 
free-free emission at cm radio wavelengths. The bolometric luminosity has contributions 
from the accretion of infalling material and nuclear burning. Up to date only a few 
systematic searches for high mass protostellar objects have been carried out 
(Molinari et al. 1996, 1998, 2000; Sridharan et al. 2002). 

We have recently started a multi-wavelength study of a sample of 18 luminous IRAS 
sources in the southern hemisphere thought to be representative of young massive 
star forming regions (Mardones, Garay, \& Bronfman 2002). The goal is to understand 
the physical and chemical differences between different stages of early evolution. 
The objects were taken from the Galaxy-wide survey of CS(2$\rightarrow$1) emission 
towards IRAS sources with IR colors typical of compact \hii\ regions (Bronfman, 
Nyman, \& May 1996). We selected sources based primarily on the observed 
CS(2$\rightarrow$1) line profiles; looking for self-absorbed lines consistent with 
inward or outward motions (e.g., Mardones 1998), and/or with extended line wings, 
possibly indicating the presence of bipolar outflows. In addition, the sources were 
required to have IRAS 100$\mu$m fluxes greater than $10^3$ Jy and to be in the 
southern hemisphere ($\delta < -20\arcdeg$). The luminosity of the IRAS sources, 
computed using the IRAS energy distribution and the distances derived by Bronfman 
(2002) are in the range $2\times10^4 - 4\times10^5$ \Lsun, implying that they contain 
at least an embedded massive star.

Most of the selected objects are expected to be associated with ultra compact (UC) 
\hii\ regions which are thought to be manifestations of newly formed massive stars 
that are still embedded in their natal molecular cloud. This expectation is confirmed 
by the radio continuum observations of Walsh et al. (1998) which show that 9 of the 
12 sources in both samples have detectable radio continuum emission (above a 
$3\sigma$ limit of 6 mJy/beam at 8.64 GHz with angular resolution of $\sim$1.5\arcsec).
The objects that were not detected at radio wavelengths are suitable candidates for 
massive stars in very early stages of evolution in which dense material is still 
falling toward a massive OB-type protostar. In this accretion phase, the high-mass 
accretion rate of the infalling material quenches the development of an UC \hii\ 
region (Yorke 1984; Walmsley 1995), and the free-free emission from the ionized 
material is undetectable at centimeter wavelengths. The mass accretion rate might 
also be large enough that the ram pressure of the infalling gas could provide the 
force to prevent the expansion of an \hii\ region. We note, however, that due to the 
limited sensitivity of the Walsh et al. (1998) survey, low-density \hii\ regions 
with emission measures smaller than $4\times10^5$ pc cm$^{-6}$ were not detectable. 
Hence, the lack of detection at the above limit does not rule out the presence of an 
optically thin compact \hii\ region. 

In this paper we report sensitive ATCA radio continuum observations toward two
sources in our sample, IRAS 15596$-$5301 and 16272$-$4837, without previously 
detected radio continuum emission to place stringent limits in their radio flux 
density. The $1\sigma$ sensitivity level of 70 $\mu$Jy at 4.8 GHz achieved in our 
observations is thirty times smaller than in previous studies and is sufficient to 
detect the emission measure corresponding to any ionizing OB star within the Galaxy. 
The main goal was to establish whether or not these objects correspond to very young 
massive objects; that is massive protostars which have not yet ionized
significant amounts of the surrounding gas. Also reported in this paper are 
millimeter continuum and molecular line observations of IRAS 15596$-$5301 and 
16272$-$4837 made with the Swedish-ESO submillimetre telescope. The latter 
observations are part of a molecular line survey toward several high-mass star 
forming regions made in order to determine their physical characteristics and 
investigate possible chemical differences.

%%%%%%%%%%%%%%%%%%%%%%%%%%%%%%%%%%%%%%%%%%%%%%%%%%%%%%%%%%%%%%%%%%%%%%%%%
% OBSERVATIONS
%%%%%%%%%%%%%%%%%%%%%%%%%%%%%%%%%%%%%%%%%%%%%%%%%%%%%%%%%%%%%%%%%%%%%%%%%

\section{Observations}

The observations were made using the Australia Telescope Compact Array 
ATCA \footnote{The Australia Telescope Compact Array is funded by the 
Commonwealth of Australia for operation as a National Facility managed by
CSIRO} in Australia, and the 15 m Swedish--ESO Submillimetre Telescope
(SEST) located on La Silla, Chile.

\subsection{ATCA}

The ATCA radio continuum observations were made in 2000 May, 23, using the 6D 
configuration, which utilizes all six antennas and covers east-west baselines 
from 77 m to 5.9 km. Observations were made simultaneously at the frequencies 
of 4.800 and 4.928 GHz, each spanning a bandwidth of 128 MHz. At these frequencies 
the FWHM primary beam of ATCA is 10\arcmin. Total integration time 
for each source was 270 minutes, obtained from 15-minute scans taken over 
a wide range of hour angles to provide good (u,v) plane coverage. The calibrator 
PKS 1600-48 was observed before and after every on-source scan in order to correct 
the amplitude and phase of the interferometer data for atmospheric and instrumental 
effects. The flux density was calibrated by observing PKS 1934-638 (3C84) for which 
values of 5.83 Jy at 4.8 GHz and 5.65 Jy at 4.9 GHz were adopted. 
Standard calibration and data reduction were performed using MIRIAD 
(Sault, Teuben, \& Wright 1995). Maps were made by Fourier transformation of the 
uniformly weighted interferometer data using the AIPS task MX. The noise level in 
each image is 70 $\mu$Jy, close to the expected theoretical limit of 50 $\mu$Jy. 
The synthesized (FWHM) beams were $2.30\arcsec\times1.68\arcsec$ for IRAS 
15596$-$5301 and $2.46\arcsec\times1.64\arcsec$ for IRAS 16272$-$4837. 

\subsection{SEST}

\subsubsection{Millimeter continuum}

The 1.2 mm continuum observations were made in 2001, September using the SIMBA 
37-channel bolometer array. The HPBW of a single element is 24\arcsec\ and 
the separation between elements on the sky is 44\arcsec. We made the observations 
in the fast mapping mode, using a scan speed of 80\arcsec\ s$^{-1}$. Our observing 
blocks consisted of 50 scan lines in azimuth of length 800\arcsec\ and separated in 
elevation by 8\arcsec, giving a map size in elevation of 400\arcsec. This block 
required $\sim$ 15 minutes of observing time. Two blocks were observed for each 
source. The data were reduced in a standard fashion, applying sky-opacity correction, 
calibration using a counts-to-flux conversion factor derived from maps of bright 
planets, baseline subtraction, and rejection of correlated sky-noise using the 
software package MOPSI. The rms noise level in the final maps is  
0.1 Jy with a pixel binning of 8\arcsec. Uncertainties in the pointing accuracy are 
estimated at 1$\sigma$ = 5\arcsec.

\subsubsection{Molecular lines}

The molecular line observations were carried out in two epochs. During June of 1999 
we used as backend the high resolution acousto-optical spectrometers 
which provided a channel separation of 43 kHz and a total bandwidth of 43 MHz. 
We observed, at the position of the IRAS sources, molecular line emission in ten 
transitions: CS(2$\rightarrow$1), CS(5$\rightarrow$4), 
CS(7$\rightarrow$6), C$^{34}$S(2$\rightarrow$1), SiO(2$\rightarrow$1), 
SO($6_5\rightarrow5_4$), HCO$^+$(1$\rightarrow$0), HCO$^+$(3$\rightarrow$2), 
H$^{13}$CO$^+(1\rightarrow$0) and C$^{18}$O(2$\rightarrow$1), with integration times 
on source ranging from 3 to 6 minutes. System temperatures were typically $\sim220$~K 
at 3 mm, $\sim350$~K at 1 mm, and $\sim700$~K at 0.8 mm. In addition we mapped, with 
angular spacings of 30\arcsec, the CS(2$\rightarrow$1) and CS(5$\rightarrow$4) 
emission within regions of 2.5\arcmin\ and the SiO(2$\rightarrow$1) emission within 
regions of 2.5\arcmin$\times1.5$\arcmin\ centered on the IRAS positions. These 
observations were made in order to investigate the kinematics of the molecular gas.

During March of 2000 we made a survey of molecular line emission using as backend 
the low resolution acousto-optical spectrometers providing a channel separation of 
700 kHz and a total bandwidth of 1.0 GHz.  The 100/150 GHz and 115/230 GHz 
pairs of SIS receivers were used to simultaneously observe lines in, respectively, 
the 3/2 mm and 3/1 mm wavelength ranges. We observed 39 spectral windows, 1.0 GHz 
wide each, within the 85 to 250 GHz frequency range. System temperatures were 
typically $\sim 200$~K at 2-mm and 3-mm, and $\sim 400$~K at 1-mm. The integration 
times on source were 5 minutes, resulting in antenna temperature rms noises of 
typically 20 mK. A 
detailed description of these observations are given in Garay et al. (2002).
All the observations were performed in the dual beam-switching mode, with a beam
separation of 11\arcmin47\arcsec\ in azimuth.

%%%%%%%%%%%%%%%%%%%%%%%%%%%%%%%%%%%%%%%%%%%%%%%%%%%%%%%%%%%%%%%%%%%%%%%%%
% RESULTS
%%%%%%%%%%%%%%%%%%%%%%%%%%%%%%%%%%%%%%%%%%%%%%%%%%%%%%%%%%%%%%%%%%%%%%%%%

\section{Results}

\subsection{Radio continuum emission.}

\begin{figure}
\epsscale{0.95}
\plotone{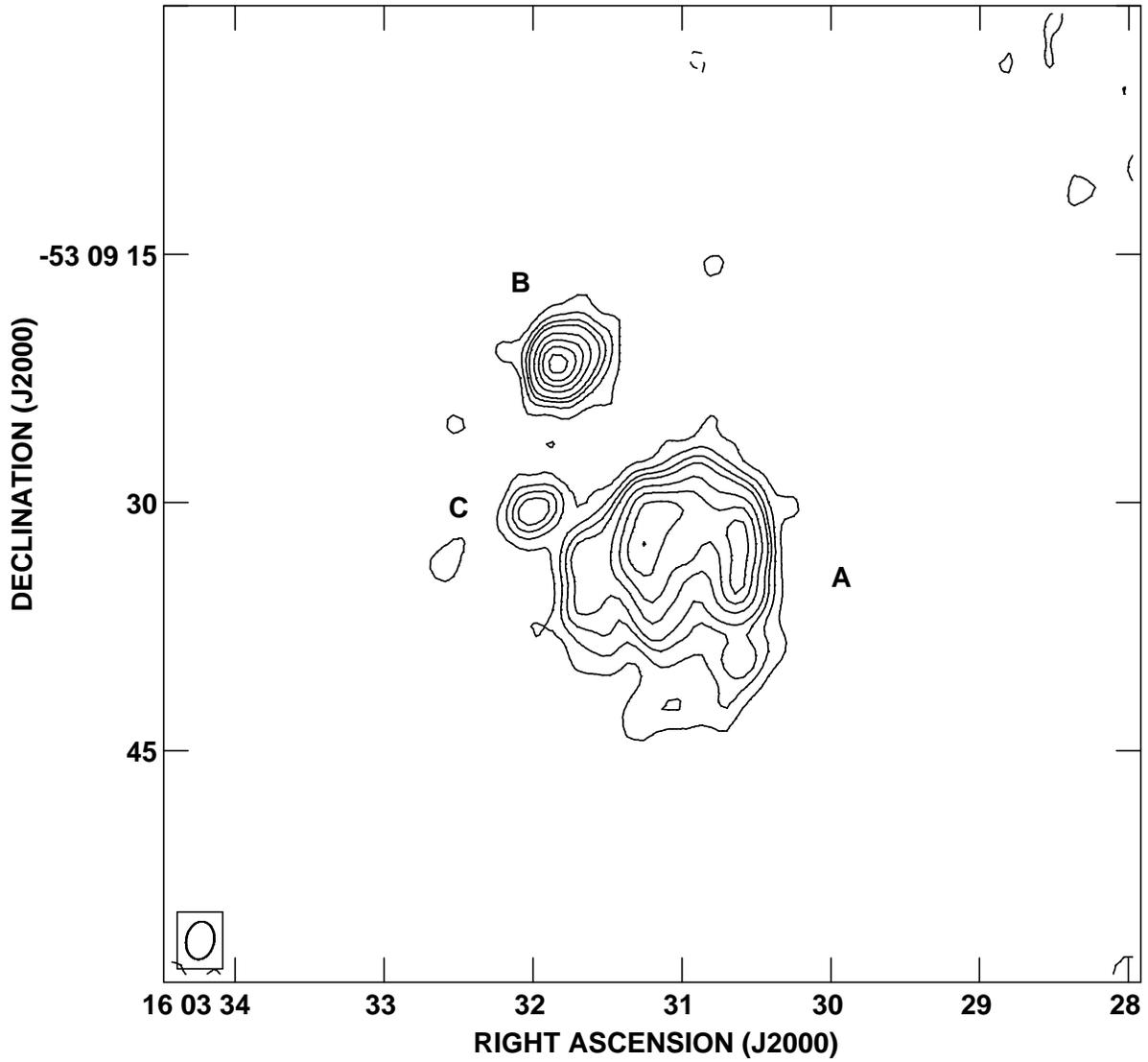}
\caption
{\baselineskip3.0pt
ATCA map of the 6-cm radio continuum emission from IRAS 15596$-$5301. The 
angular resolution is $2.30\arcsec\times1.68\arcsec$ (FWHM beam is shown 
in the lower left corner). Contour levels are -3, 3, 6, 9, 12, 18, 24, 30, and 36 times 
70 $\mu$Jy beam$^{-1}$ (= $1\sigma$ noise level).
\label{fig-radiomap}}
\end{figure}

Figure~\ref{fig-radiomap} shows a map of the 6 cm radio continuum emission from 
IRAS 15596$-$5301. Three compact radio sources (labeled A, B, and C) were detected 
within a region of 30\arcsec\ in diameter. Their positions, flux densities, and 
sizes are given in Table 1. Their peak flux densities are in the range 1.1 to 2.8 
mJy/beam, close to the sensitivity limit in the survey of Walsh et al. (1998). 
The less compact source (object A) appears to have an irregular shell-like 
structure.

No radio continuum emission was detected toward 16272-$4837$ within a region of 
3\arcmin\ in radius centered on the IRAS position. We set a new 3$\sigma$ upper limit 
for the 6 cm flux density of 16272-$4837$ of 0.2 mJy, about thirty times smaller than 
that previously reported by Walsh et al. (1998). We detected, however, 
emission from a region located about 4\arcmin\ northwest of the array phase center, 
with a total flux density (corrected by primary beam) of 122 mJy. The morphology of 
the emission is complex, showing an elongated bipolar-like structure with a peak 
located at $\alpha_{2000}=
16^{\rm h}30^{\rm m}35\fs0$, $\delta_{2000}=-48\arcdeg40\arcmin47.6\arcsec$. 

\subsection{Millimeter continuum emission.}

\begin{figure}
\epsscale{0.75}
\plotone{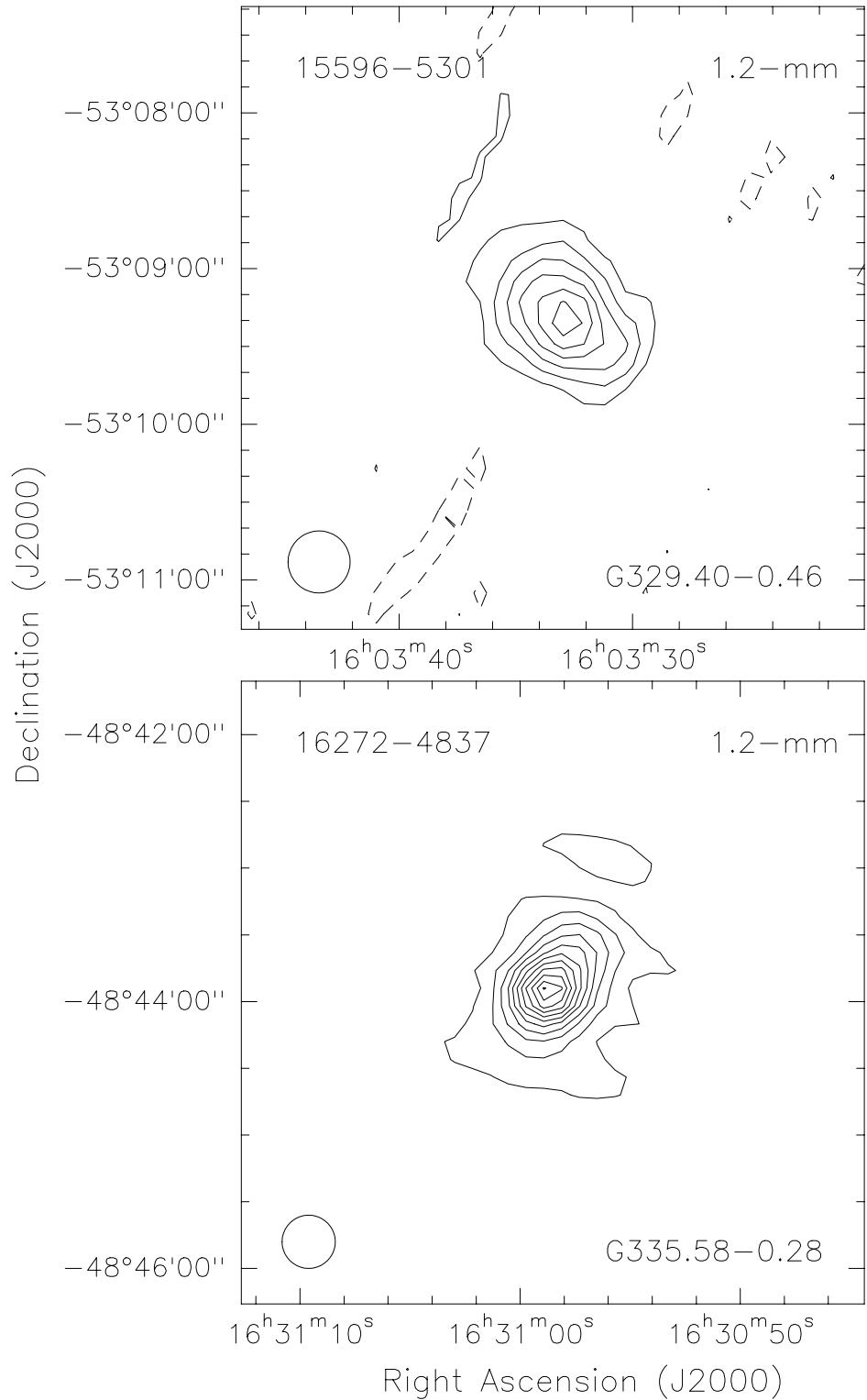}
\caption
{\baselineskip3.0pt
Maps of 1.2-mm continuum emission observed with SEST. The angular resolution 
is 24\arcsec (FWHM beam is shown in the lower left corner). Top: IRAS 15596$-$5301. 
Contour levels are -1, 1, 2, 3, 4, 5, and 6 times 0.3 Jy beam$^{-1}$. Bottom: IRAS 
16272$-$4837. Contour levels are -1, 1, 2, 3, 4, 5, 6, 7, 8, 9 and 10 times 0.5 
Jy beam$^{-1}$. 
\label{fig-mmmap}}
\end{figure}

Figure~\ref{fig-mmmap} presents maps of the 1.2-mm continuum emission
observed with SEST, revealing the presence of strong emission towards both IRAS sources. 
The 1.2-mm emission associated with IRAS 15596$-$5301 has a peak position at 
$\alpha_{2000} = 16^{\rm h}03^{\rm m}32\fs9$, $\delta_{2000} = 
-53\arcdeg09\arcmin20\arcsec$, a flux density of $5.8\pm0.3$ Jy and a peak flux 
density of 2.0 Jy/beam. It shows an elongated morphology, with deconvolved major and 
minor axes of 42.1\arcsec\ and 24.6\arcsec\ at P.A. 46\arcdeg. The 1.2-mm emission 
detected towards IRAS 16272$-$4837 has a peak position at $\alpha_{2000}=
16^{\rm h}30^{\rm m}58\fs7$, $\delta_{2000}=-48\arcdeg43\arcmin55\arcsec$, a flux 
density of $13.8\pm0.9$ Jy and a peak flux density of 4.6 Jy/beam. It has an elongated 
morphology, with deconvolved major and minor axes of 41.0\arcsec\ and 24.7\arcsec\ at 
P.A. 142\arcdeg.  

\subsection{Molecular line emission.}

\begin{figure}
\epsscale{0.85}
\plotone{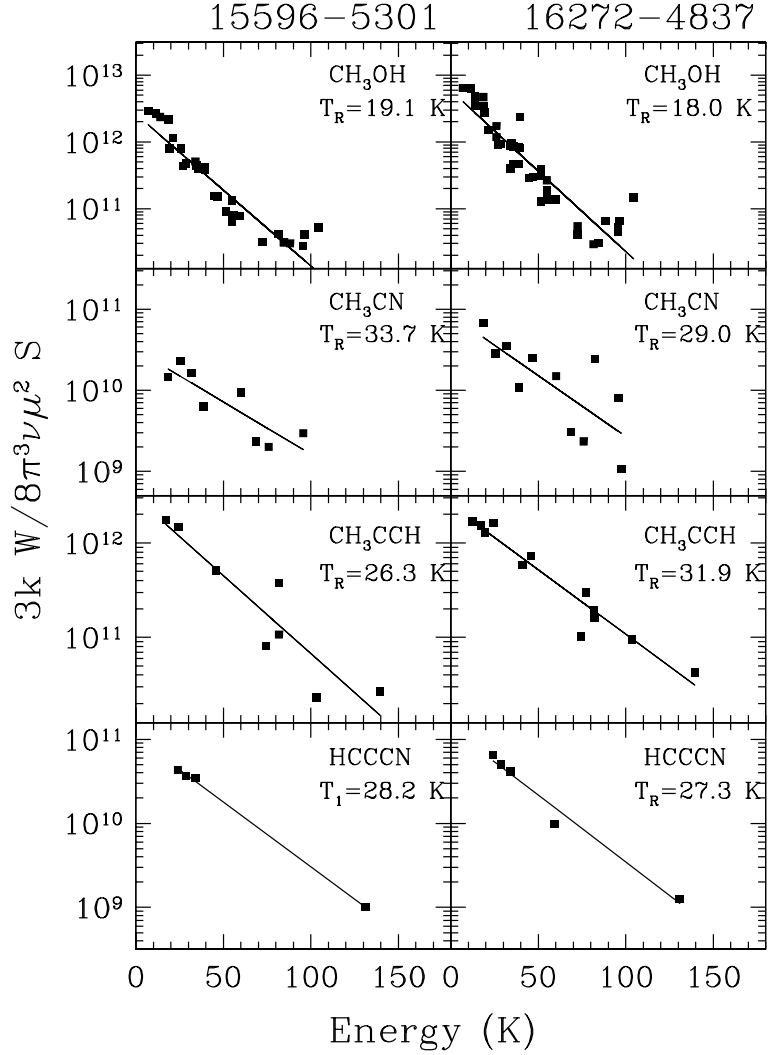}
\caption
{\baselineskip3.0pt
Rotational diagrams of the emission from selected species
observed toward the IRAS 15596$-$5301 and 16272$-$4837 massive dense cores.
From top to bottom: CH$_3$OH, CH$_3$CN, CH$_3$CCH, and HCCCN.
The lines correspond to least squares linear fits to the observed data. The 
derived rotational temperature is given in the upper right corner.
\label{fig-trotdiag}}
\end{figure}

The molecular line survey indicate that IRAS 15596$-$5301 and 16272$-$4837 have 
similar chemical characteristics. This is illustrated in Figure~\ref{fig-trotdiag} 
which shows rotational diagrams of the emission from selected species (CH$_3$OH, 
CH$_3$CN, CH$_3$CCH, and HCCCN) with optically thin lines and at least five 
observed transitions. The average rotational temperature derived from these 
diagrams is 27 K for both sources. Overall we detected emission from more than 
20 species toward both objects. Table 2 summarizes the derived properties of the 
molecular emission in selected species. Column densities, rotational temperatures, 
and abundances relative to CO are given in columns 2-4. The column density of 
optically thick species, such as CO and CS, were determined as follows. The optical 
depths of a pair of an optically thick line and an optically thin line of an 
isotopic species are determined, for a given abundance ratio, from the ratio of 
their observed brightness temperatures. From these opacities and assuming an 
excitation temperature of 30 K we then computed the total column density assuming 
that energy levels are populated according to local thermodynamic equilibrium (see 
Bourke et al. 1997 for a description of the method). For the derivation of CO column 
densities we used the optically thick $^{13}$CO($1\rightarrow0$) and optically thin 
C$^{18}$O($1\rightarrow0$) pair of lines, whereas for CS we used the 
CS($2\rightarrow1$) and C$^{34}$S($2\rightarrow1$) pair of lines. For species with 
optically thick lines it is also possible to determine the filling factor of the 
emission within the observed beam. Filling factors of $\sim$0.5 for $^{13}$CO and 
$\sim$0.13 for CS were estimated for both sources. 

\begin{figure}
\epsscale{0.70}
\plotone{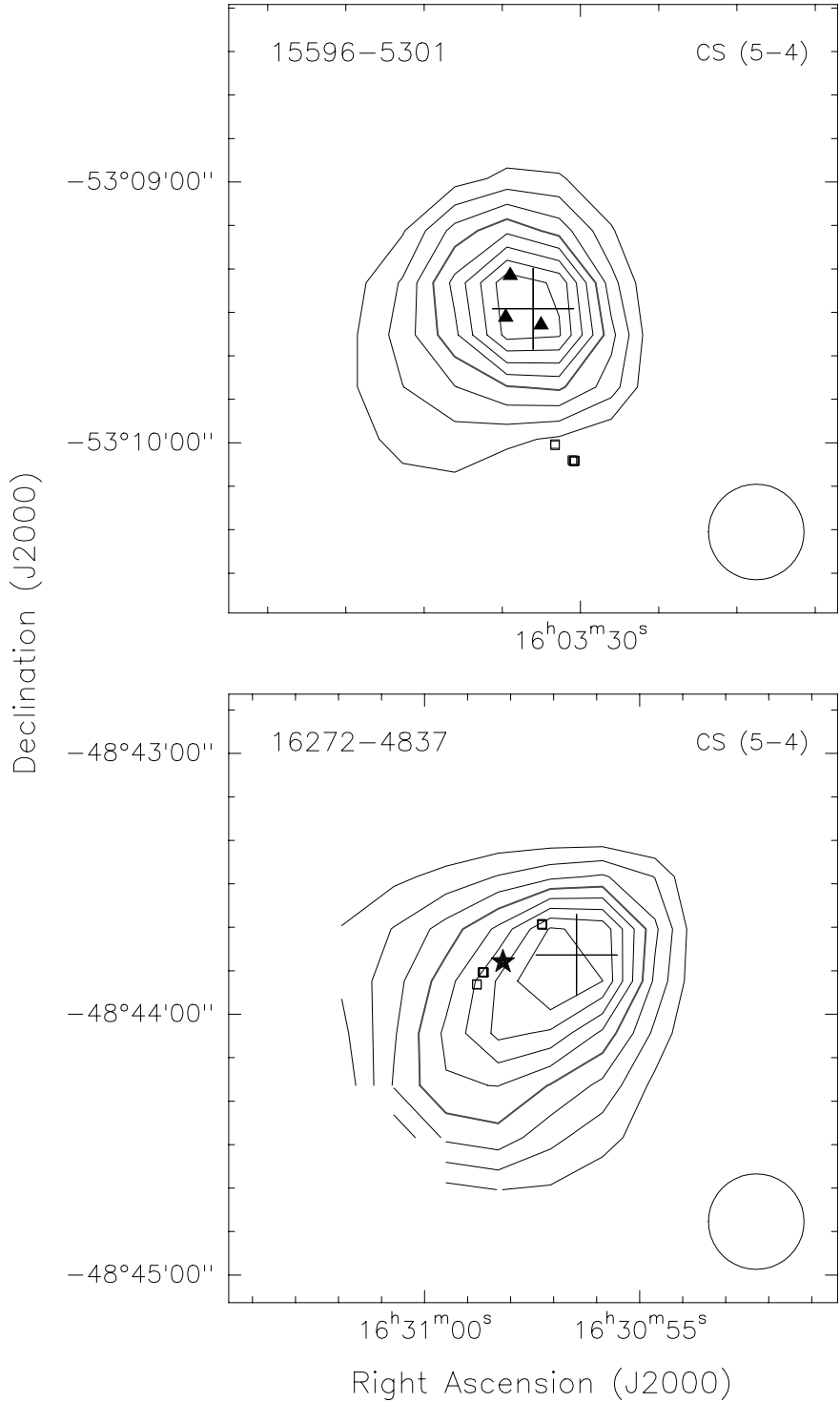}
\caption
{\baselineskip3.0pt
Maps of CS(5$\rightarrow$4) line emission observed with SEST. The angular 
resolution is 22\arcsec (FWHM beam is shown in the lower right corner).
Top: IRAS 15596$-$5301. Contour levels are 20, 30, 40, 50, 60, 70, 80, and 90 \% 
of the peak velocity integrated antenna temperature of 11.4 K \kms. The triangles 
mark the positions of the compact \hii\ regions. Bottom: IRAS 16272$-$4837. Contour 
levels are 20, 30, 40, 50, 60, 70, 80, and 90 \% of the peak velocity integrated 
antenna temperature of 7.2 K \kms. The star marks the peak position of the 22$\mu$m 
MSX source.  In both panels the cross mark the position of the IRAS source and 
the squares the position of 6.67 GHz methanol masers (Walsh et al. 1998).
\label{fig-cs54maps}}
\end{figure}

Figure~\ref{fig-cs54maps} shows maps of the CS($5\rightarrow4$) line emission
from both sources. The morphology of the dense gas from IRAS 15596$-$5301
is roughly circular, with a mean angular diameter of 37\arcsec. This implies a 
cloud radius of 0.41 pc at a distance of 4.6 kpc. The spatial distribution of the 
dense gas from IRAS 16272$-$4837 is elongated, with major and minor axes of 
61\arcsec\ and 42\arcsec, respectively. This implies radii of 0.50 and 0.35 pc at 
a distance of 3.4 kpc. From the observed CS size and derived CS column density we 
estimate that the CS density is $4\times10^{-3}$ cm$^{-3}$ in IRAS 15596$-$5301 and 
$2\times10^{-3}$ cm$^{-3}$ in 16272$-$4837. Further, assuming a [CS/H$_2$] abundance 
ratio of $1\times10^{-8}$ (cf. van der Tak et al. 2000), we find that the molecular 
hydrogen density and the mass of molecular gas are $4\times10^{5}$ cm$^{-3}$ and 
$6.6\times10^3$ \Msun\ for IRAS 15596$-$5301, and $2\times10^{5}$ cm$^{-3}$ and 
$2.6\times10^3$ \Msun\ for IRAS 16272$-$4837. Thus, the molecular observations 
show that both sources are associated with dense, massive molecular cores. The mass 
of the cores can, alternatively, be estimated assuming that they are in virial 
equilibrium (MacLaren et al. 1988). From the observed size and average line width 
in the CS(5-4) line we derive virial masses of $2.3\times10^{3}$ \Msun\ for IRAS 
15596$-$5301 ($R=0.41$ pc; $\Delta v$=5.17 \kms) and $2.6\times10^{3}$ \Msun\ for 
IRAS 16272$-$4837 ($R=0.42$ pc; $\Delta v$=5.44 \kms). 

\begin{figure}
\epsscale{0.95}
\plotone{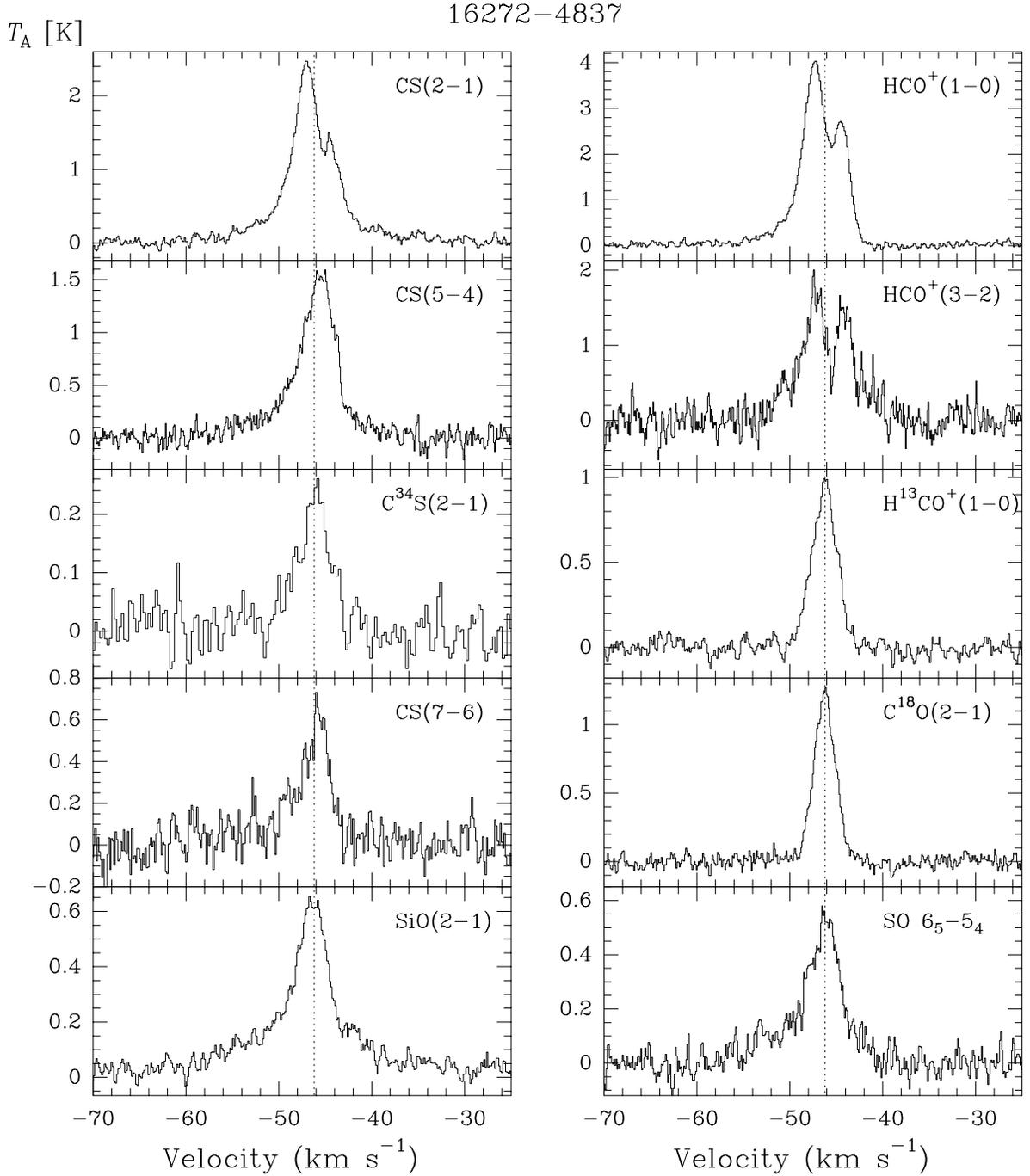}
\caption
{\baselineskip3.0pt
Spectra of the emission in several molecular lines observed toward the peak 
position of the IRAS 16272$-$4837 massive core.  
Transitions are given in the upper right corner of the spectra. The vertical
dotted line indicates the systemic velocity of the ambient gas of $-46.2$ \kms.
\label{fig-specpeak16272}}
\end{figure}

Figure~\ref{fig-specpeak16272} shows the spectra observed, with the high 
resolution spectrometer, toward the peak position of the IRAS 16272$-$4837 core. The 
CS(2$\rightarrow$1), HCO$^+$(1$\rightarrow$0) and HCO$^+$(3$\rightarrow$2) spectra 
show double-peaked line profiles, with a bright blue-shifted peak at the velocity 
of $\sim-47.1$ \kms\ and a weaker red-shifted peak at the velocity of $\sim-44.6$ 
\kms. On the other hand, the profiles of the C$^{34}$S(2$\rightarrow$1) and 
H$^{13}$CO$^+$(1$\rightarrow$0) lines show a symmetric single component with a peak 
line center velocity of $\sim-46.0$ \kms. These spectroscopic signatures suggest 
that the bulk of the molecular gas toward IRAS 16272$-$4837 is undergoing 
large-scale inward motions. Infalling motions traced by optically thick molecular 
lines are expected to produce line profiles showing blue asymmetry, whereas optically 
thin lines are expected to exhibit symmetrical profiles (Mardones et al. 1997).

\begin{figure}
\epsscale{0.75}
\plotone{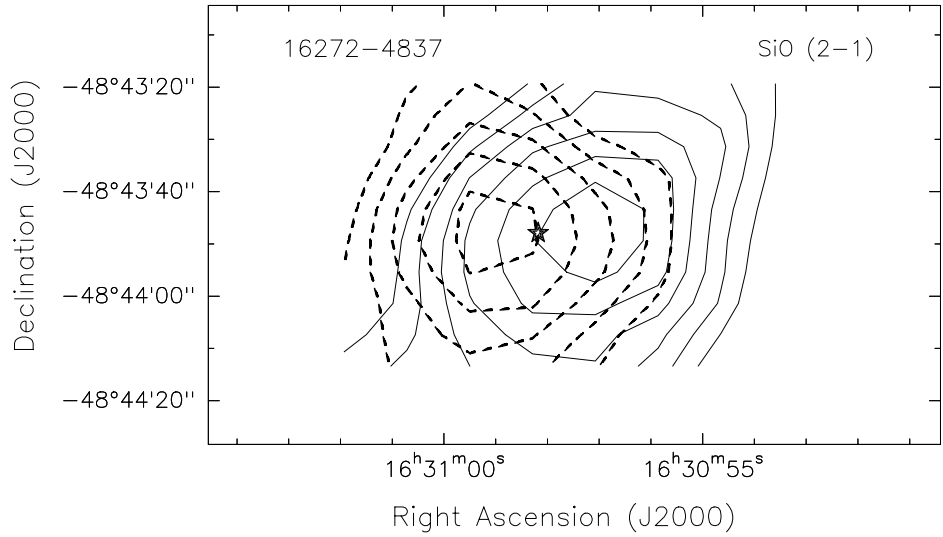}
\caption
{\baselineskip3.0pt
Contour maps of velocity integrated SiO($2\rightarrow1$) line wing emission from 
IRAS 16272$-$4837. Continuous lines represent blue-shifted emission
($v_{flow}: -19.5$ to $-3.8$ \kms) and dashed lines red-shifted emission ($v_{flow}:
4.2$  to 15.9 \kms). The star marks the position of the 22$\mu$m MSX source. 
Contour levels are at 0.60, 0.72, 0.84, 0.96, 1.08 and 1.20 K \kms.
\label{fig-siooutflow}}
\end{figure}

In some species, particularly SiO and SO, the spectra show the presence of strong 
wing emission indicative of outflow activity. In the SiO($2\rightarrow1$) line the 
wing emission extends up to velocities of $-19.6$ \kms\ and 19.5 \kms\ relative 
to the ambient cloud velocity of $-46.2$ \kms. Figure~\ref{fig-siooutflow} shows 
contour maps of the velocity integrated SiO emission in the range of flow velocities 
from $-19.5$ to $-3.8$ \kms\ (continuous lines; blueshifted gas) and from $4.2$ to 
$15.9$ \kms\ (dotted lines; redshifted gas). The flow velocity is defined as 
$v_{flow} = v_{\small LSR}-v_o$, where $v_o$ is the velocity of the emission from 
quiescent ambient gas. The spatial distribution of the wing emission clearly shows a 
bipolar structure with redshifted and blueshifted lobes extending in opposite 
directions of a 22$\mu$m source detected by MSX (see later discussion). The 
bipolar structure shows a moderate degree of collimation, with the blueshifted 
emission seen mainly toward the west and the redshifted emission seen mainly toward 
the east, suggesting that the underlying wind is a wide angle wind. 

\subsection{Mid-infrared emission.}

\begin{figure}
\epsscale{0.95}
\plotone{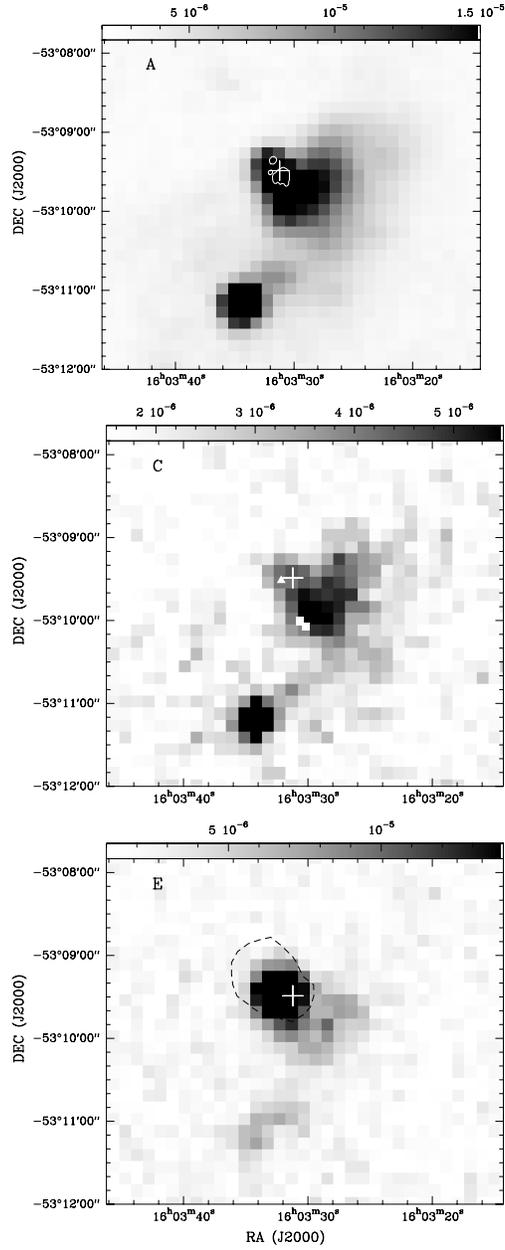}
\caption
{\baselineskip3.0pt
MSX images (gray-scale) towards IRAS 15596$-$5301. The cross marks the peak 
position of the IRAS source. Top: A-band image. The contour delineates the radio 
continuum emission at the level of 0.5 mJy. Middle: C-band image. The squares and 
triangles mark the position of 6.67 GHz methanol and hydroxyl masers, respectively.
Bottom: E-band image. The dotted contour indicates the 1.2-mm dust emission at 
the level of 0.5 Jy. 
\label{fig-msxace15596}}.
\end{figure}

We have used the database from the {\sl Midcourse Space Experiment} (MSX; Price 
1995) survey of the Galactic plane (Egan et al. 1998) to investigate the mid-infrared 
emission, in the four MSX broad bands (A:6.8-10.8 $\mu$m, C:11.1-13.2 $\mu$m, 
D:13.5-15.9 $\mu$m, and E:18.2-25.1 $\mu$m), from both IRAS sources.
Figure~\ref{fig-msxace15596} presents grey scale images of the emission in the A, C, 
and E bands of a region of about 4\arcmin\ toward IRAS 15596$-$5301. These images 
show the presence of two sources within the central 1\arcmin\ region: a compact 
object, brighter in the E band image, and an extended object, brighter in the C band, 
with a peak located about 30\arcsec\ southwest of the former. Both sources are seen 
in emission in the four MSX bands. The three radio continuum sources detected within 
this region (delineated by the 0.5 mJy contour level in the upper panel of 
Figure~\ref{fig-msxace15596}), as well as the peak position of IRAS 15596$-$5301 
(marked by the cross), lie projected within the face the compact object. Further, 
the extent of the 1.2-mm dust continuum emission (delineated by the 0.5 Jy contour 
level in the lower panel of Figure~\ref{fig-msxace15596}) is similar to that of the 
E-band infrared emission. These results show that the compact MSX object is 
intimately associated with the 15596$-$5301 massive molecular core. We measured flux 
densities of 5.4, 4.1, and 24.3 Jy in the C, D, and E bands, respectively. 
The extended object shows strong emission in the A and C bands, which are sensitive 
to emission from polycyclic aromatic hydrocarbon (PAH) features (7.7 and 8.3 $\mu$m 
in A band; 11.3 and 12.6 $\mu$m in C band). Since no radio emission is detected 
toward this MSX source, we suggest that it might correspond to a photodissociated 
region excited by a non-ionizing star. 

IRAS 16272$-$4837 is associated with an infrared dark cloud seen in absorption against 
bright mid-infrared emission in the A, C, and D bands. This is shown in the top 
two panels of Figure~\ref{fig-msxace16272} which presents grey scale images of the 
emission in the A, C, and E bands toward IRAS 16272$-$4837. There is a good agreement 
between the molecular, millimeter and mid-infrared morphologies, as illustrated by the 
contour of the 1.2-mm continuum emission shown in Figure~\ref{fig-msxace16272}. From 
the molecular hydrogen column density derived from the line observations (of 
$3.4\times10^{23}$ cm$^{-2}$), assuming a gas-to-dust mass ratio of 110 (Draine \& Lee 
1984) and a dust opacity at 8$\mu$m of $10^3$ cm$^{2}$ gr$^{-1}$ (Ossenkopf \& Henning 
1994), we estimate that the optical depth at 8$\mu$m is $\sim12$, indicating that this 
cloud is optically thick at MID-IR. Further, the MSX observations in the E-band 
show the presence, near the center of the infrared dark cloud, of an object seen in 
emission located at $\alpha_{2000} = 16^{\rm h}30^{\rm m}58\fs18$, $\delta_{2000} =
-48\arcdeg43\arcmin47.9\arcsec$ (see bottom panel of Figure~\ref{fig-msxace16272}).
We finally note that the MSX object seen near the top right corner of 
Figure~\ref{fig-msxace16272}, which is clearly extended in the A-band image,
is associated with the IRAS object 16269-4834 and coincident with the radio source 
detected $\sim$4\arcmin\ northwest of 16272$-$4837. 

%%%%%%%%%%%%%%%%%%%%%%%%%%%%%%%%%%%%%%%%%%%%%%%%%%%%%%%%%%%%%%%%%%%%%%%%%
% DISCUSSION
%%%%%%%%%%%%%%%%%%%%%%%%%%%%%%%%%%%%%%%%%%%%%%%%%%%%%%%%%%%%%%%%%%%%%%%%%

\section{Discussion}

\subsection{Spectral energy distribution}
 
\begin{figure}
\epsscale{0.90}
\plotone{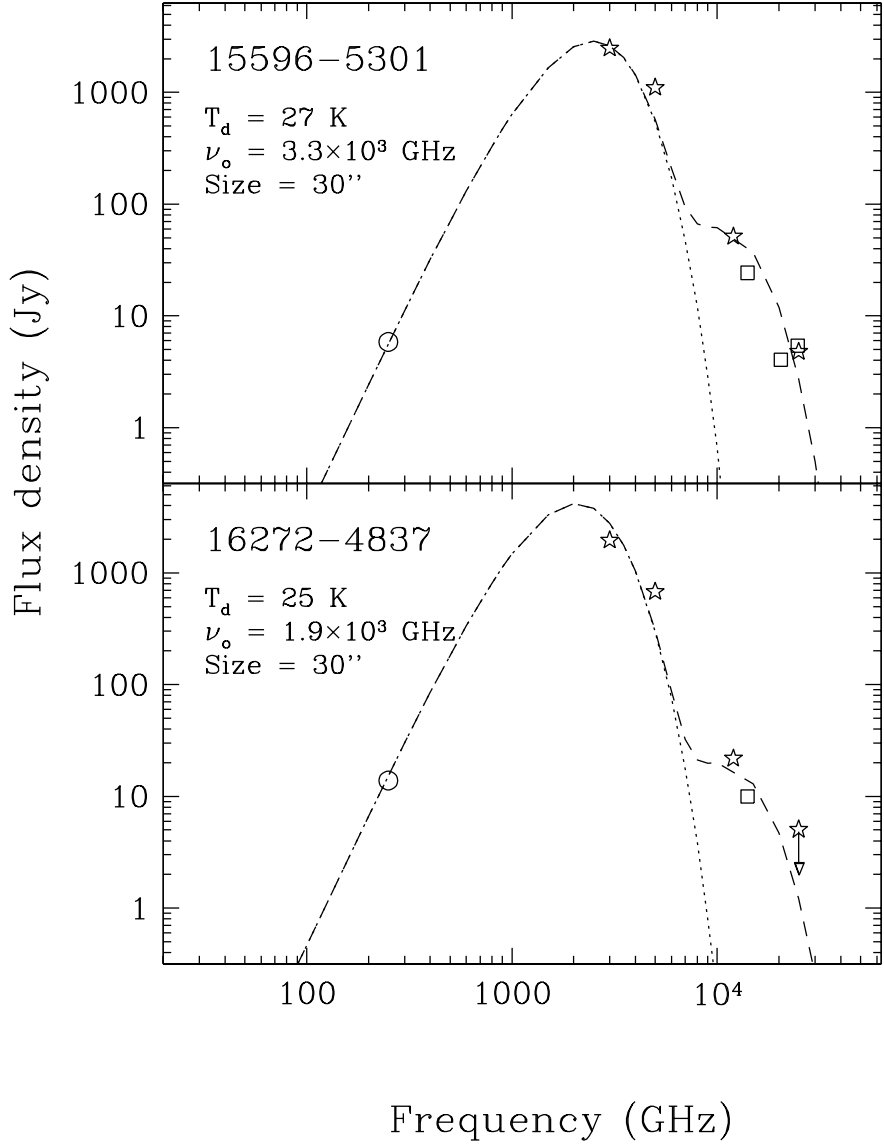}
\caption
{\baselineskip3.0pt
Spectral energy distributions. 
Stars mark IRAS fluxes, squares MSX fluxes, and the circle the SIMBA flux.  
The long-dashed curve is a fit to the spectrum using two modified blackbody functions 
of the form $B_{\nu}(T_d) \left[ 1 - \exp(-(\nu/\nu_o)^{\beta}) \right]$, with 
different temperatures. The short-dashed line indicates the fit for the colder 
temperature component (fit parameters indicated on the upper left). 
Top: IRAS 15596$-$5301. Bottom: IRAS 16272$-$4837 (12$\mu$m flux is an upper limit). 
\label{fig-sed}}
\end{figure}

Figure~\ref{fig-sed} shows the spectral energy distribution (SED) of IRAS 
15596$-$5301 and 16272$-$4837 from 12 $\mu$m to 1.2 mm, which is mainly due to 
thermal dust emission. We fitted the SED with modified blackbody functions of 
the form $ B_{\nu}(T_d)\left[1-\exp(-\tau_{\nu})\right]\Omega_s~, $
where $\tau_{\nu}$ is the dust optical depth, $B_{\nu}(T_d)$ is the Planck 
function at the dust temperature $T_d$, and $\Omega_s$ is the solid angle subtended 
by the dust emitting region. The opacity was assumed to vary with frequency as
$\nu^{\beta}$, i.e. $\tau_{\nu}= \left(\nu/\nu_o\right)^{\beta}$, where
$\nu_o$ is the frequency at which the optical depth is unity. Due to the limited 
number of spectral points we have set the value of $\beta$ equal to 2.0, consistent 
with tabulated opacities (Ossenkopf \& Henning 1994) and derived values for high 
mass star forming regions (Molinari et al. 2000). A single temperature model 
produced poor fits, underestimating the emission observed at wavelengths smaller 
than 25$\mu$m, and therefore we used a model with two temperature components.

From the fits (long-dashed lines) we derive that the colder dust component 
(short-dashed lines) toward 15596$-$5301 and 16272$-$4837 have, respectively, 
temperatures of 27 and 25 K, angular sizes (assuming a Gaussian flux 
distribution) of 30\arcsec (FWHM), and wavelengths at which the opacity is unity of 
$\sim$90 $\mu$m and 140 $\mu$m. The temperature of the hot dust component
is 100 K for 15596$-$5301 and 115 K for 16272$-$4837. The thermal dust emission at 
1.2 mm is therefore 
optically thin ($\tau_{1.2mm}\sim 5\times10^{-3}$), and thus the observed flux 
density at 1.2 mm allows to obtain an additional mass estimate 
of the dense cores. In general, for an isothermal dust source the total gas mass,
$M_{g}$, is given in terms of the observed flux density, $S_{\nu}$, at an
optically thin frequency, $\nu$, by (e.g. Chini, Krugel, \& Wargau 1987)
$$ M_{g} = {{S_{\nu} D^2}\over{R_{dg} \kappa_{\nu} B_{\nu}(T_d)}} ~~, $$
where $\kappa_{\nu}$ is the mass absorption coefficient of dust, $R_{dg}$ is
the dust-to-gas mass ratio (assuming 10\% He), and $B_{\nu}(T_d)$ is the Planck 
function at the dust temperature $T_d$. The main source of uncertainty in the 
conversion of the observed flux density into gas mass is the $R_{dg}\kappa_{\nu}$ 
factor, or total mass opacity, which is a poorly known quantity (e.g. Gordon 1995).
Using a dust opacity at 1.2 mm of 1 cm$^2$ g$^{-1}$, as computed by Ossenkopf 
\& Henning (1994) for dense and cold protostellar cores, $R_{dg}=0.01$, the fitted 
dust temperatures, and the observed flux densities, we derive masses of $1.4\times10^3$ 
\Msun\ for IRAS 15596$-$5301 and $2.0\times10^3$ \Msun\ for IRAS 16272$-$4837.
These masses derived from the dust emission are in good agreement with those 
derived from the molecular line intensities and from the virial assumption.

\subsection{Evolutionary stages } 

\subsubsection{IRAS 15596$-$5301 (G329.40-0.46)}

The radio continuum observations toward IRAS 15596$-$5301 indicate the presence, in 
a region of $\sim$0.3 pc in radius, of three distinct \hii\ regions with diameters 
ranging from 0.06 to 0.2 pc. The multiple structure of the ionized gas is typical 
of galactic \hii\ regions, and is most likely due to the presence of a cluster of 
exciting stars. If components A, B, and C are excited by individual ZAMS stars, the 
rate of UV photons needed to ionize them (see Table 3) imply exciting stars with 
spectral types of B0, B0.5, and B1, respectively. The total luminosity emitted by 
this cluster of B stars, as inferred from the radio observations, is 4.1$\times10^4$ 
\Lsun. On the other hand, the total far-infrared luminosity computed using the 
IRAS fluxes (see Casoli et al. 1986) is $\sim6.5\times10^4$ \Lsun\ (assuming a 
distance of 4.6 kpc; Bronfman 2002). The difference between the radio derived 
luminosity and the IRAS luminosity could be explained by the presence of dust 
within the \hii\ regions. Garay et al. (1993) found that the fraction of Lyman 
continuum photons absorbed by dust within \hii\ regions is typically~55\%. 
Alternatively, it could be explained by the presence, in addition to the B stars, 
of several less massive stars that will contribute to the FIR luminosity but that 
are not hot enough to contribute to ionization. It is not easy, however, to 
disentangle which of these effects play the predominant role.

The compact \hii\ regions are found projected toward the peak 
of the CS($5\rightarrow4$) emission map (see Figure~\ref{fig-cs54maps}), 
suggesting that they are deeply embedded within the dense molecular core.  
From the observed sizes, and assuming a sound speed in the ionized gas of 11.4 
\kms, we estimate that the \hii\ regions have dynamical ages between 3$\times 10^3$ 
to 8$\times 10^3$ yrs. If these correspond to the actual ages of the 
compact \hii\ regions, then we should conclude that they are very young objects. 
The dynamical time-scales, however, may not provide a realistic estimate of the 
actual age of \hii\ regions. The large number of UC \hii\ regions and their 
short dynamical ages poses the well known problem that the rate of massive star 
formation appears to be much greater than other indicators suggest (Wood \& 
Churchwell 1989, Churchwell 1990). Due to the high density of the molecular gas 
in which they are embedded, we suggest instead that the \hii\ regions within the 
IRAS 15596$-$5301 massive core might be in pressure equilibrium with the surrounding 
dense ambient medium, and are currently stalled at their equilibrium radius. (e.g. 
De Pree, Rodr\'\i guez, \& Goss 1995). The molecular density of the ambient gas 
needed to stall an \hii\ region at radius, $R_f$, is given by (e.g. Garay \& Lizano 
1999)   
$$ \left({{n_{H_2}}\over{10^{5}~\rm cm^{-3}}}\right) =  1.9
\left({{N_{\rm u}}\over{10^{49}~\rm s^{-1}}}\right)^{1/2}
\left({{T_e}\over{10^4~\rm K}}\right)
\left({{30~\rm K}\over{T_o}}\right)
\left({{\rm pc}\over{R_f}}\right)^{3/2}~~, $$
where $N_u$ is the rate of ionizing UV photons emitted by the exciting star,
$T_e$ is the electron temperature of the ionized gas, and $T_o$ is the temperature 
of the ambient gas. Using the observed radius and the derived ionizing rate of 
UV photons of the \hii\ regions within IRAS 15596$-$5301, we find that molecular 
densities of $\sim5\times10^5$ cm$^{-3}$ are needed for them to be pressure 
confined by the dense environment. These densities are similar to those derived 
from the molecular observations. The time needed for the \hii\ regions to achieve 
pressure equilibrium are between $1.0\times10^5$ yrs to $2.5\times10^5$ yrs, implying 
that massive star formation started within this core more than $2.5\times10^5$ yrs
ago. We conclude that the dense massive core is in an advanced stage of early 
evolution, in which multiple OB star formation have already taken place near its 
central region.

\subsubsection{IRAS 16272$-$4837 (G335.58-0.28)}

The total far-infrared luminosity of IRAS 16272$-$4837 computed using the IRAS fluxes 
(see Casoli et al. 1986) is $\sim2.4\times10^4$ \Lsun\ (assuming a distance of 3.4 
kpc; Bronfman 2002). The luminosity obtained integrating under the fitted curve in 
Figure~\ref{fig-sed} is similar to the IRAS luminosity. 
The high luminosity suggests that the IRAS 16272$-$4837 massive core hosts a
young massive protostar inside; and the lack of radio emission suggests that it
is still undergoing an intense accretion phase. Models of 
massive envelopes accreting onto a young massive central B type star (e.g. Osorio et 
al. 2000) require accretion rates in the envelopes $\dot M >5\times10^{-4} M_{\odot}$ 
yr$^{-1}$ in order to fit the observed SEDs. The high-mass accretion rate of the 
infalling material quenches the development of an UC \hii\ region (Yorke 1984), and the 
free-free emission from the ionized material is undetectable at centimeter wavelengths. 

This hypothesis is supported by the characteristics of the line profiles observed 
toward IRAS 16272$-$4837, which suggest that the molecular gas is undergoing infalling 
motions. From the spectra of the optically thick HCO$^+(1\rightarrow0)$ line we 
measure a velocity difference between the red and blue peaks of 2.7 \kms, 
and brightness temperature of the blue peak, red peak, and dip of 5.4, 3.6, and 2.9 K, 
respectively. From the spectra of the optically thin H$^{13}$CO$^+(1\rightarrow0)$ 
line we measure a FWHM line width of 3.18 \kms. From these values, using the simple 
model of contracting clouds of Myers et al. (1996), we derive a characteristic inward 
speed of 0.5 \kms. We note that this value is considerably smaller than the free-fall 
velocity expected for a cloud with a total mass of $\sim3\times10^3$\Msun\ at its 
outer envelope radius of 0.4 pc, suggesting that the collapse is not dynamical. Using 
the derived values of the infall speed, molecular density, and core size, we obtain a 
mass infall rate, $\dot{M}_{in}$, of $1\times10^{-2} M_{\odot}$ yr$^{-1}$, large 
enough to prevent the development of an UC \hii\ region. The high value of the mass 
infall rate rises the question as to which is the fraction of the total luminosity 
due to accretion.  The accretion luminosity, $L_{acc}$, is 
$$ L_{acc} = {{G~ f~ \dot{M}_{in}~ M_p}\over{R_p}}~~,$$
where $f$ is the fraction of the large scale mass infall rate that goes into accretion 
onto the protostar, $M_p$ is the mass of the protostar, and $R_p$ the radius where 
the shock occurs. None of these three parameters are known for IRAS 16272$-$4837. 
Assuming $M_p\sim10$\Msun, $R_p\sim3\times10^{12}$ cm, and $f\sim0.05$ (e.g, 
Norberg \& Maeder 2000) we obtain $L_{acc}\sim3.6\times10^3$ \Lsun, about 15\% of 
the total luminosity. We emphasize that this value of the accretion luminosity 
corresponds only to a rough estimate, particularly because the value of $f$ is 
highly uncertain.

Additional evidence for IRAS 16272$-$4837 to be in a collapsing stage is provided 
by the presence of bipolar outflowing gas, phenomenon which is thought to be closely 
related to accretion processes. The 22$\mu$m object lies at the center of symmetry 
of the SiO outflow, suggesting it is intimately associated with the energy source of 
the outflow. The early evolutionary stage of this region is also sustained by the 
presence of 6.67 GHz methanol masers, which are thought to be signposts of young 
regions of massive star formation (Walsh et al. 1997, 1998). Notice that the 22$\mu$m 
source is elongated, with the 6.67 GHz methanol masers being aligned along its major 
axis.  Finally, we mention that the high value of the mass to luminosity 
ratio of IRAS 16272$-$4837, $M/L = 0.083$, about 4 times higher than that of IRAS 
15596$-$5301, is another indicator of its youth, as argued by Sridharan et al (2002).

%%%%%%%%%%%%%%%%%%%%%%%%%%%%%%%%%%%%%%%%%%%%%%%%%%%%%%%%%%%%%%%%%%%%%%%%%
% CONCLUSIONS
%%%%%%%%%%%%%%%%%%%%%%%%%%%%%%%%%%%%%%%%%%%%%%%%%%%%%%%%%%%%%%%%%%%%%%%%%

\section{Summary}

We undertook sensitive radio continuum observations at 4.8 GHz, using ATCA, 
and 1.2 millimeter continuum and molecular line observations, using SEST, 
toward IRAS 15596$-$5301 and 16272$-$4837, two luminous objects 
(${\cal L}>2\times10^4$\Lsun) thought to represent massive star forming regions 
in early stages of evolution.  The main results and conclusions are summarized 
as follows: 

1. The 4.8 GHz radio continuum observations toward IRAS 15596$-$5301 show the 
presence of three distinct compact sources with diameters of 0.06 to 0.2 pc, all 
located within a region of 30\arcsec\ in diameter. If they are regions of 
ionized gas, they have electron densities of $8\times10^2 - 2\times10^3$ cm$^{-3}$ 
and are excited by early B type stars. The 1.2-mm continuum observations show that 
the dust emission arises from an elongated region with major and minor axes of 
42.1\arcsec and 24.6\arcsec, respectively. The observed 1.2-mm flux density, of 5.8 
Jy, implies a mass of $1.4\times10^3$ \Msun. The line observations indicate a region 
of molecular gas with a radius of $\sim0.4$ pc, a molecular hydrogen density of 
$4\times10^5$ cm$^{-3}$, and a rotational temperature of $\sim27$ K. We conclude that 
IRAS 15596$-$5301 corresponds to a dense massive ($\sim3\times10^3$ \Msun) molecular 
core, with a radius of about 0.4 pc, that is in the compact \hii\ region phase of 
evolution, already hosting a cluster of massive B type stars. Further, we suggest 
that the regions of ionized gas excited by these stars are in pressure equilibrium 
with the ambient molecular gas.

2. No radio continuum emission at 4.8 GHz was detected from IRAS 16272$-$4837 up to 
a $3\sigma$ limit of 0.2 mJy. The 1.2-mm continuum observations show strong dust 
emission arising from a region with an elongated morphology, with major and minor 
axes of 41.0\arcsec\ and 24.7\arcsec.  The observed flux density at 1.2-mm, of 
$13.8\pm0.9$ Jy, implies a mass of $2.0\times10^3$ \Msun. The line observations 
indicate a molecular gas region with a size of $\sim0.4$ pc, a molecular hydrogen 
density of $\sim2\times10^5$ cm$^{-3}$, and a rotational temperature of $\sim27$ K.  
We suggest that this object corresponds to a dense massive core in a very early 
evolutionary stage, distinguished by being luminous but not associated with an UC 
\hii\ region (eg. Cesaroni et al. 1994; Hunter et al. 1998; Molinari et al. 1998). 
The core is being heated by a recently formed massive star embedded at its center 
which is still accreting at high rates, the main heating agent being the accretion 
luminosity, and have not yet produced a detectable ultra compact \hii\ region. As such, 
this will become a key object for the study at submillimeter wavelengths with high 
angular resolution of the earliest stages of the formation of massive stars, and their 
related accretion and outflow processes.

\acknowledgements

G.G. and D.M. gratefully acknowledge support from the Chilean {\sl Centro de 
Astrof\'\i sica} FONDAP No. 15010003.  M.G.B. acknowledges support from the 
Australian Research Council. 

\vfill\eject

%%%%%%%%%%%%%%%%%%%%%%%%%%%%%%%%%%%%%%%%%%%%%%%%%%%%%%%%%%%%%%%%%%%%%%%%%
% REFERENCES
%%%%%%%%%%%%%%%%%%%%%%%%%%%%%%%%%%%%%%%%%%%%%%%%%%%%%%%%%%%%%%%%%%%%%%%%%

\newcommand\rmaap   {RMA\&A~}
\newcommand\rmaacs   {RMA\&A Conf. Ser.~}

\clearpage

%%%%%%%%%%%%%%%%%%%%%%%%%%%%%%%%%%%%%%%%%%%%%%%%%%%%%%%%%%%%%%%%%%%%%%%%%
% TABLES
%%%%%%%%%%%%%%%%%%%%%%%%%%%%%%%%%%%%%%%%%%%%%%%%%%%%%%%%%%%%%%%%%%%%%%%%%
\newpage
\clearpage

\begin{deluxetable}{cllcc}
\tablewidth{0pt}
\tablecaption{OBSERVED PARAMETERS OF RADIO CONTINUUM EMISSION \label{tbl-obs}}
\tablehead{
\colhead{\hii\ region}      & \multicolumn{2}{c}{Position}  &
\colhead{Flux density}      & \colhead{Angular size}    \\
\cline{2-3}
\colhead{}                  & \colhead{$\alpha$(2000)}      &
\colhead{$\delta$(2000)}         & \colhead{(mJy)}          &
\colhead{(\arcsec$\times$\arcsec)} \\
}
\startdata
 A & $16^{\rm h} 03^{\rm m} 31{\rlap.}{^s}01$ & $-53\arcdeg\ 09\arcmin\
 32{\rlap.}{\arcsec}9$
   &  43.4  & 9.7$\times$8.0 \\
 B & 16~03~~31.80   & $-53~~09~~21.5$ & 9.0 & 3.0$\times$2.5 \\
 C & 16~03~~31.91   & $-53~~09~~31.1$ & 3.0 & 6.1$\times$2.6 \\
\enddata
\end{deluxetable}

\begin{deluxetable}{lcrccccc}
%\tabletypesize{\small}
\tablewidth{0pt}
\tablecaption{MOLECULAR GAS PARAMETERS \label{tbl-molpar}}
\tablehead{
\colhead{Molecule} & \multicolumn{3}{c}{IRAS 15596} & \colhead{} &
   \multicolumn{3}{c}{IRAS 16272-4837} \\
\cline{2-4} \cline{6-8}
\colhead{} & \colhead{$T_R$} & \colhead{N} & \colhead{$[X/CO]$} & &
  \colhead{$T_R$} & \colhead{N} & \colhead{$[X/CO]$} \\
\colhead{} & \colhead{(K)} & \colhead{(cm$^{-2})$} & \colhead{} &
 \colhead{} & \colhead{(K)} & \colhead{(cm$^{-2})$} & \colhead{} \\
}
\startdata

CO        & 30\tablenotemark{a} & $8.1\times10^{19}$ & 1 & ~ &
  30\tablenotemark{a} & $2.1\times10^{19}$ & 1 \\
$^{13}$CO  & 30\tablenotemark{a} & $1.5\times10^{18}$ & $1.9\times10^{-2}$ & ~ &
  30\tablenotemark{a} & $3.8\times10^{17}$ & $1.8\times10^{-2}$ \\
C$^{18}$O  & 30\tablenotemark{a} & $1.1\times10^{17}$ & $1.4\times10^{-3}$ & ~ &
  30\tablenotemark{a} & $2.7\times10^{16}$ & $1.3\times10^{-3}$ \\
CS        & 30\tablenotemark{a}  & $9.8\times10^{15}$ & $1.2\times10^{-4}$ & ~ &
  30\tablenotemark{a} & $4.2\times10^{15}$ & $2.0\times10^{-4}$ \\
C$^{34}$S & 30\tablenotemark{a}  & $4.4\times10^{14}$ & $5.4\times10^{-6}$ & ~ &
  30\tablenotemark{a} & $1.9\times10^{14}$ & $9.0\times10^{-6}$ \\
H$_2$CO  & --- & --- & --- & ~ &
  27.0 & $9.4\times10^{13}$ & $4.5\times10^{-6}$ \\
CH$_3$OH  & 19.1 & $2.7\times10^{14}$ & $3.3\times10^{-6}$ & ~ &
  18.0 & $5.5\times10^{14}$ & $2.6\times10^{-5}$ \\
CH$_3$CCH & 26.3 & $2.1\times10^{14}$ & $2.6\times10^{-6}$ & &
  31.9 & $2.3\times10^{14}$ & $1.1\times10^{-5}$ \\
CH$_3$CN & 33.7 & $9.0\times10^{12}$ & $1.1\times10^{-7}$ & &
  29.0 & $1.9\times10^{13}$ & $9.0\times10^{-7}$ \\
CH$_3$CHO & 25.0 & $7.2\times10^{12}$ & $8.9\times10^{-8}$ & &
  21.0 & $1.3\times10^{13}$ & $6.2\times10^{-7}$ \\
SiO       & 10.4 & $2.9\times10^{12}$ & $3.6\times10^{-8}$ & &
   9.9 & $1.0\times10^{13}$ & $4.8\times10^{-7}$ \\
SO        & 23.0 & $6.3\times10^{13}$ & $7.8\times10^{-7}$ & &
  11.7 & $8.5\times10^{13}$ & $4.0\times10^{-6}$ \\
HCCCN  & 28.2 & $4.1\times10^{13}$ & $5.1\times10^{-7}$ & &
  27.3 & $5.1\times10^{13}$ & $2.4\times10^{-6}$ \\
OCS    & 37.5 & $3.9\times10^{13}$ & $4.8\times10^{-7}$ & &
  28.5 & $1.3\times10^{14}$ & $6.2\times10^{-6}$ \\
SO$_2$ & 27.6 & $1.9\times10^{13}$ & $2.3\times10^{-7}$ & &
  25.9 & $1.6\times10^{13}$ & $7.6\times10^{-7}$ \\
\enddata
\tablenotetext{a}{Assumed excitation temperature.}
\end{deluxetable}

\begin{deluxetable}{ccrrrl}
\tablewidth{0pt}
\tablecaption{DERIVED PARAMETERS OF \hii\ REGIONS \label{tbl-obs}}
\tablehead{
\colhead{\hii\ region}  & \colhead{Diameter}  &
\colhead{EM}     & \colhead{$n_e$}  & \colhead{N$_u$}  & \colhead{S.T.}   \\
\colhead{}       & \colhead{(pc)} & \colhead{(pc cm$^{-6}$)}  &
\colhead{(cm$^{-3}$)} & \colhead{(s$^{-1}$)}    & \colhead{} \\
}
\startdata
 A & 0.196 & $2.4\times10^5$ & $9.2\times10^2$ & $8.3\times10^{46}$ & B0 \\
 B & 0.061 & $5.2\times10^5$ & $2.4\times10^3$ & $1.7\times10^{46}$ & B0.5 \\
 C & 0.089 & $8.0\times10^4$ & $7.8\times10^2$ & $5.7\times10^{45}$ & B1 \\
\enddata
\end{deluxetable}

\vfill\eject

%%%%%%%%%%%%%%%%%%%%%%%%%%%%%%%%%%%%%%%%%%%%%%%%%%%%%%%%%%%%%%%%%%%%%%%%%
% FIGURES
%%%%%%%%%%%%%%%%%%%%%%%%%%%%%%%%%%%%%%%%%%%%%%%%%%%%%%%%%%%%%%%%%%%%%%%%%


\clearpage

\clearpage

\clearpage

\clearpage

\clearpage

\clearpage

\clearpage

\clearpage

\begin{figure}
\epsscale{0.95}
\plotone{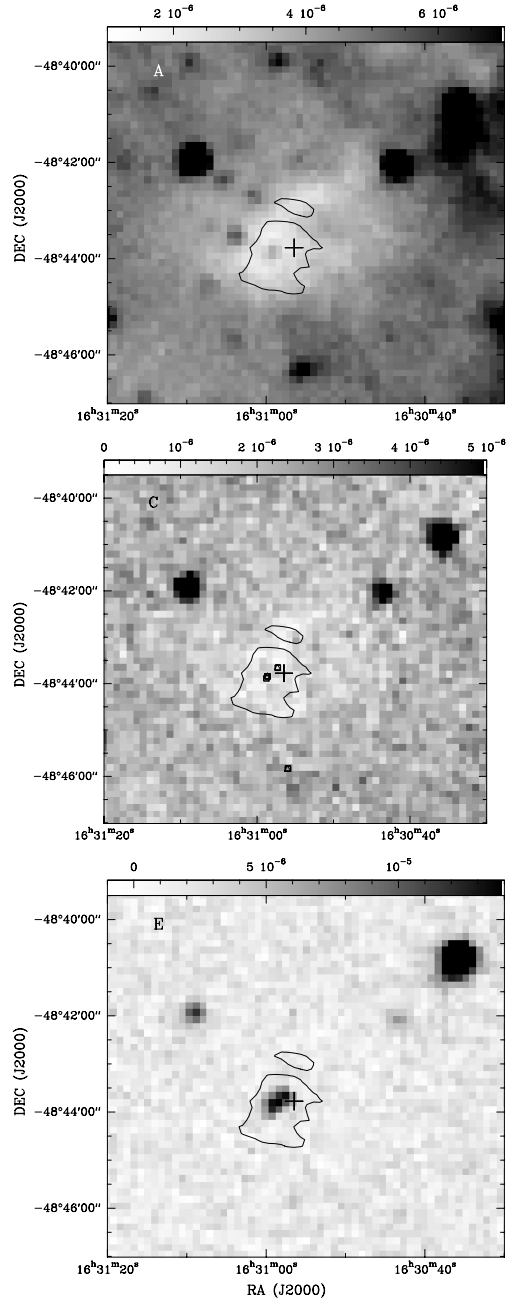}
\caption
{\baselineskip3.0pt
MSX images (gray-scale) towards IRAS 16272$-$4837. The cross marks the peak 
position of the IRAS source and the contour delineates the 1.2-mm continuum 
emission at the level of 0.5 Jy. Top: A-band image. The light areas in the
gray-scale correspond to regions of lower emission than the bright Galactic
plane emission, and indicates the infrared extinction cloud associated with IRAS
16272$-$4837. Middle: C-band image. The squares and triangles mark the position of
6.67 GHz methanol and hydroxyl masers, respectively.
Bottom: E-band image. 
\label{fig-msxace16272}}
\end{figure}

\clearpage

\label{fig-msxace16272}

\clearpage

\end{document}